 \documentstyle[aps,prl,epsf]{revtex}

\begin{document}

\draft

%%%
 \twocolumn[{
%%%
 \widetext

\title{
Phase diagram of a Heisenberg spin-Peierls model with quantum phonons
}

\author{
Robert J. Bursill,\cite{email} Ross H. McKenzie and Chris J. Hamer
}

\address{
School of Physics, University of New South Wales, Sydney, NSW 2052, 
Australia.
}

\date{Received 24 December, 1998}

\maketitle
\mediumtext

\begin{abstract}
Using a new version of the density-matrix renormalization group we
determine the phase diagram of a model of an antiferromagnetic
Heisenberg spin chain where the spins interact with quantum phonons.
A quantum phase transition from a gapless spin-fluid state to a gapped
dimerized phase occurs at a non-zero value of the spin-phonon coupling.
The transition is in the same universality class as that of a
frustrated spin chain, to which the model maps in the anti-adiabatic
limit. We argue that realistic modeling of known spin-Peierls
materials should include the effects of quantum phonons.
\end{abstract}

%63.20.-e Phonons in crystal lattices
%63.20.Kr Phonon-electron and phonon-phonon interactions
%63.20.Ls Phonon interactions with other quasiparticles
%63.22.+m Phonons in low-dimensional structures and small particles
%63.70.+h Statistical mechanics of lattice vibrations and displacive
%71.10.Hf Non-Fermi-liquid ground states, electron phase diagrams and
%         phase transitions in model systems
%71.10.Pm Fermions in reduced dimensions (anyons, composite fermions,
%         Luttinger liquid, etc.)
%71.20.Ps Other inorganic compounds
%71.20.Rv Polymers and organic compounds
%71.30.+h Metal-insulator transitions and other electronic transitions
%71.38.+i Polarons and electron-phonon interactions (see also 63.20.K)
%         phase transitions
%71.45.Lr Charge-density-wave systems
%75.10.Jm Quantized spin models
%75.30.Et Exchange and superexchange interactions
%75.30.Kz Magnetic phase boundaries (including magnetic transitions,
%         metamagnetism, etc.)
%75.40.-s Critical-point effects, specific heats, short-range order
%75.40.Cx Static properties (order parameter, static susceptibility,
%         heat capacities, critical exponents, etc.)
%75.40.Mg Numerical simulation studies
%75.50.-y Studies of specific magnetic materials
%75.50.Ee Antiferromagnetics

%%%
 }]
%%%
 \narrowtext

	Challenged by the discovery of high-temperature
superconductivity in doped antiferromagnets, our 
understanding of quantum magnetism in low dimensions
has increased significantly over the past decade \cite{auerbach}.
However, the effect of the interaction of 
quantum spin systems with
further degrees of freedom such as disorder,
phonons, and holes produced by doping is
still poorly understood.
Interest in models of spins interacting with phonons
has increased significantly since the discovery
of a spin-Peierls transition in the inorganic
compound CuGeO$_3$ \cite{hase}.
The availability of large, high-quality 
single crystals has led to much more extensive experimental
studies \cite{review} than                    
on the organic spin-Peierls materials studied in the 1970's \cite{bray}.

 The fact that a spin-1/2 antiferromagnetic Heisenberg chain
is unstable to a {\it static} uniform dimerization \cite{bray,cross}
is known as the spin-Peierls instability.
This occurs because dimerization
opens a gap $\Delta$ in the spin excitation
spectrum and lowers the total magnetic energy
at a greater rate than
the increase in elastic energy due to the dimerization.
Until very recently
almost all theoretical treatments have used this {\it static}
picture which
we can only expect to be valid in the adiabatic regime where
typical frequencies $\omega$ of the phonons associated
with the dimerization are much smaller than the magnetic
energy scales such as $\Delta$ and the antiferromagnetic
exchange integral $J$.
It has recently been pointed out that CuGeO$_3$ is not in this
adiabatic regime \cite{uhrig,gros,wellein},
stimulating several numerical studies with
dynamical phonons \cite{wellein,augier}.

\begin{figure}[htbp]
%
%%%
 \centerline{\epsfxsize=8.4cm \epsfbox{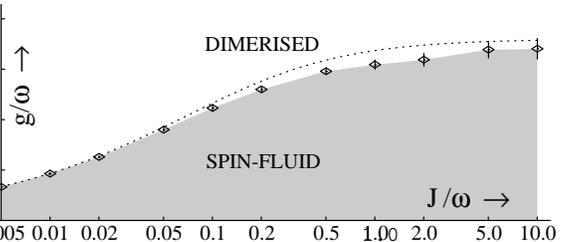}}
%%% \vskip 0.5cm
\caption{
Zero temperature phase diagram of the spin-Peierls antiferromagnetic 
chain of spins interacting with quantum phonons (Eqn.\ 
(\protect\ref{H})). For small spin-phonon coupling $g$ the system is a 
gapless spin-fluid. For large $g$ the system is dimerized and has an 
energy gap. The diamonds with error bars denote the phase boundary from 
this DMRG study. The dotted line is (Eqn.\ (\protect\ref{gc})) the phase 
boundary which results from an approximate mapping onto the $J_1$-$J_2$ 
model (frustrated antiferromagnetic chain, Eqn.\ (\protect\ref{j1j2})) 
which becomes exact in the anti-adiabatic limit $J / \omega \rightarrow 
0$.
}
\label{phase_diagram}
\end{figure}

In this Letter we study a model of
a spin-1/2 antiferromagnetic Heisenberg chain
interacting with quantum phonons using a powerful
new numerical technique that allows an essentially
exact treatment of {\it both} the spins and the phonons
at a fully quantum-mechanical level.
Our main result is the phase diagram in Fig.\ \ref{phase_diagram}
in which the adiabaticity parameter
$J / \omega$ varies over several decades.
We find that the spin-phonon coupling must be
larger than some  non-zero critical value
for the spin-Peierls instability to occur.                 
This is in contrast to the static case ($\omega/J \to 0$) for which
dimerization occurs for any value of the coupling.
Hence, quantum lattice fluctuations can destroy
Heisenberg spin-Peierls order.
We find that the quantum phase transition from the spin-fluid
state to the gapped state is in the same universality
class as the dimerization transition of the
$J_1$-$J_2$ frustrated spin chain.
Our results have important implications for the 
modeling of spin-Peierls materials.

The model we study is one of the simplest possible.
It consists of a local phonon on each site
and the antiferromagnetic exchange on neighbouring sites
varies linearly with the difference between the phonon amplitudes
on the two sites. The Hamiltonian is
\begin{equation}
{\cal H} = \sum_{i=1}^N
  \left( J + g \left( b_{i+1}+ b_{i+1}^\dagger - b_i
-b_i^\dagger  \right) \right)
 \vec{S}_i \cdot \vec{S}_{i+1} + \omega \sum_{i=1}^N b_i^\dagger b_i.
\label{H}
\end{equation} 
Here $\vec{S}_i$ is the $S = 1/2$ spin operator on site $i$ and
 $b_i$ destroys a phonon of frequency $\omega$ on site $i$.
 We assume a periodic chain of $N$ sites.

Insight into this model can be obtained by considering
the anti-adiabatic limit ($\omega >> J$).
One can then integrate out the phonon degrees of
freedom to obtain the following effective
Hamiltonian for the spin degrees of freedom \cite{fuk}
\begin{equation}
{\cal H}_{\text{eff}}  =  
J_1 \sum_{i=1}^N \vec{S}_i \cdot \vec{S}_{i+1}
+ J_2 \sum_{i=1}^N  \vec{S}_i \cdot \vec{S}_{i+2}
\label{j1j2}
\end{equation}
 where $J_1 = J  + g^2 / \omega$
and $J_2  =   g^2 / 2 \omega $.
Uhrig \cite{uhrig} recently obtained the same Hamiltonian, calculating
 $J_1$ and $J_2$ to next order in $J/\omega$.
\begin{eqnarray}
J_1 & = & J + g^2 / \omega - 3 g^2 J / 2 \omega^2 + \ldots
\label{j1}
\\
J_2 & = & g^2 / 2 \omega + 3 g^2 J / 2 \omega^2 + \ldots.
\label{j2}
\end{eqnarray}
The frustrated spin chain Eqn.\ (\ref{j1j2}) or $J_1$-$J_2$ model
has been extensively studied and is well understood.
If $\alpha \equiv J_2 / J_1$ then at a critical value of 
$\alpha = \alpha_{\text{c}} = 0.241167(5)$ 
the model undergoes a quantum phase transition
from a gapless spin-fluid state with
quasi-long-range antiferromagnetic order to
a gapped phase with long-range dimer order \cite{affleck,nomura1}.
Uhrig pointed out that this 
implies that in the anti-adiabatic regime (\ref{H}) possesses a
{\em non-zero} critical coupling $g_{\text{c}}$.
To  second order in $J/\omega$,
\begin{equation}
g_{\text{c}}^2 / \omega =
\frac{\alpha_{\text{c}} J}
{1/2 - \alpha_{\text{c}} + 3 ( 1 + \alpha_{\text{c}} )J / 2 \omega}.
\label{gc}
\end{equation}
We have confirmed this result numerically (see Fig.\ 
\ref{phase_diagram}).
Furthermore, this non-zero critical coupling $g_{\text{c}}$
still occurs well into the adiabatic regime. It is interesting that 
although (\ref{gc}) is only valid to second order in $J / \omega$ it 
gives a good description of $g_{\text{c}}$ up to $J / \omega \sim 1$.

Models such as (\ref{H}), which 
involve bosons are a challenge to                                    
study numerically due to the large number of degrees of freedom
per site. The density matrix renormalization group (DMRG) method
\cite{white} has the potential for obtaining definitive results
for these models by studying very large systems.  
Several schemes based on the DMRG have recently
been developed to treat models 
involving phonons \cite{caron,schemes,bursill}.
We employ a new ``four-block'' DMRG method \cite{bursill} which 
allows us to treat the phonons and spins on an equal footing
and to study systems as large as 256 sites.
This is in contrast to some recent exact diagonalization
studies of spin-phonon models that were limited
to small systems and/or used uncontrolled truncations of
the phonon degrees of freedom \cite{wellein,augier}.
We previously used this method to obtain the phase diagram
of the Holstein model with spinless fermions \cite{holstein}.

The four-block method can be used to calculate the ground state
energy $E_0$ and the singlet and triplet gaps $\Delta_{\text{ss}}$
and $\Delta_{\text{st}}$ for periodic systems \cite{bursill}. Table
\ref{dmrg_convergence} shows the DMRG convergence of the gaps with
the {\em single} truncation parameter $\epsilon$ \cite{bursillnote}
for a representative parameter set. It can be seen that the gaps are
sufficiently well resolved to be useful for finite-size scaling
analyses. The error of around 0.1\% in the $N = 128$ site system
is typical of the error in the largest systems studied for a given
set of parameters.

We determine the critical coupling using the gap-crossing method used 
by Okamoto and Nomura \cite{nomura1} to determine the critical
 coupling $\alpha_{\text{c}}$ in the frustrated Heisenberg model 
(\ref{j1j2}).
 The convergence of the crossover coupling $\alpha_{\text{c}}(N)$ with
 $N$ is rapid due to the absence of logarithmic
corrections at the critical point \cite{affleck,nomura1,nomura2}.
If the system is gapless with quasi-long-range N\'{e}el order
for $0 \leq g \leq g_{\text{c}}$,
 the lowest excitation is the triplet state, i.e.\
$\Delta_{\text{st}} < \Delta_{\text{ss}}$ (for sufficiently large $N$)
 and $\Delta_{\text{st}}, \Delta_{\text{ss}} \rightarrow 0$ as
$N \rightarrow \infty$.
If for $g > g_{\text{c}}$ the system has a non-zero gap $\Delta$
 and is dimerised with a doubly degenerate ground state,
then the first excited singlet state becomes
 degenerate with the ground state in the bulk limit \cite{nomura2}.
 That is, $\Delta_{\text{ss}} < \Delta_{\text{st}}$ (for sufficiently
large $N$), $\Delta_{\text{ss}} \rightarrow 0$, and
$\Delta_{\text{st}} \rightarrow \Delta > 0$ as $N \rightarrow \infty$.
A finite lattice crossover coupling $g_{\text{c}}(N)$ is defined by
$\Delta_{\text{st}} = \Delta_{\text{ss}}$. As shown in Table 
\ref{gc_convergence}, $g_{\text{c}}(N)$ rapidly approaches a limit as 
$N \rightarrow \infty$. This limit is the critical coupling 
$g_{\text{c}}$ separating gapless and gapped phases. For the 
$J/\omega > 1$ cases, where the $N$ dependence is substantial, 
$g_{\text{c}}(N)$ is well described by the functional form 
$g_{\text{c}}(N) \sim g_{\text{c}} - A \exp(- B N)$
and non-linear fitting is used to determine $g_{\text{c}}$ 
\cite{fitting_note}. The resulting phase boundary is plotted in Fig.\ 
\ref{phase_diagram}. The DMRG, discretization and fitting errors in 
$g_{\text{c}}$ are estimated to be no greater than a few percent.

 From conformal invariance the finite-size energies of the spin-fluid 
should satisfy \cite{nomura1}:
\begin{eqnarray}
E_0 & \sim & N \epsilon_\infty + \frac{\pi v_0}{6 N} + \ldots,
\label{v0}
\\
\frac{1}{4}
\left( 3 \Delta_{\text{st}} + \Delta_{\text{ss}} \right)
& \sim & \frac{\pi v_1}{N} \left( 1 + \ldots \right),
\label{v1}
\end{eqnarray}
where $\epsilon_\infty$ is the bulk ground state energy density and
$ v_0 = v_1 = v_\sigma $ is the spin wave velocity.
The combination of the gaps in Eqn.\ (\ref{v1}) is chosen
to cancel the logarithmic corrections.

\begin{figure}[htbp]
%
%%%
 \centerline{\epsfxsize=8.4cm \epsfbox{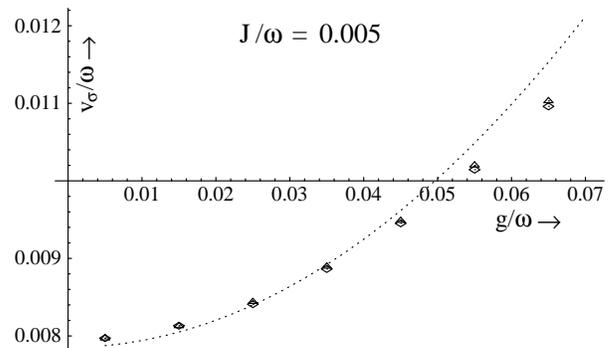}}
%%% \vskip 0.5cm
%
\caption{
The spin wave velocity $v_\sigma$ as a function of the
electron-phonon coupling $g$ for a large phonon frequency
$J / \omega = 0.005$ (anti-adiabatic regime). DMRG results (from
extrapolating the $N = 16$ and 32 data using (\protect\ref{v0})) are
indicated by diamonds. The triangles are obtained by solving the 
$N = 16$ and 32 site $J_1$-$J_2$ model (Eqn.\ (\protect\ref{j1j2})), 
with $J_1$ and $J_2$ given by (\protect\ref{j1}) and (\protect\ref{j2}). 
The dotted line arises from applying an approximate result for 
$v_\sigma$ \protect\cite{fledderjohann} to the same $J_1$-$J_2$ model.
}
\label{spin_wave}
\end{figure}

We have performed a number of consistency checks on our results.
First, in Fig.\ \ref{spin_wave}, $v_\sigma $ as determined by
(\ref{v0}) is plotted as a function of $g$ for a phonon frequency
deep into the anti-adiabatic regime $(J / \omega = 0.005)$, together
with the same quantity determined from the corresponding $J_1$-$J_2$
model. The results from the two approaches agree well.
This confirms the mapping between the two models in the
anti-adiabatic regime.
Second, we note that the DMRG results for the phase boundary
 agree well with the result (\ref{gc}) from the mapping
 in the anti-adiabatic limit (See the dotted line in Fig.\
\ref{phase_diagram}).
Third, for general phonon frequencies, we calculate
the ratio $v_0 / v_1$
which should equal unity.
At $g = g_{\text{c}}$ it is one 
within errors expected from corrections to scaling and DMRG truncation,
over the range of frequencies studied. 
Values vary from $0.98 \pm 0.04$ for $J / \omega = 0.005$ to
$1.07 \pm 0.10$ for $J / \omega = 10$.

\begin{figure}[htbp]
%
%%%
 \centerline{\epsfxsize=8.4cm \epsfbox{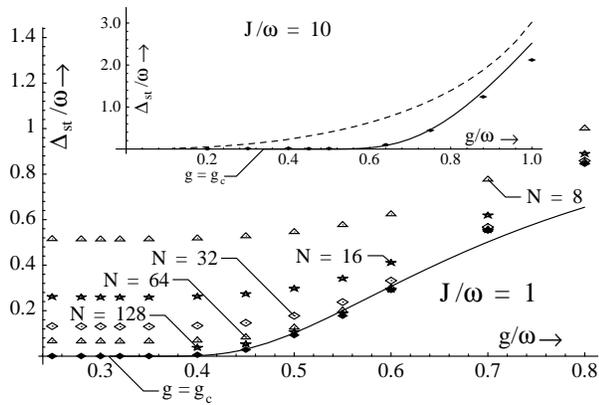}}
%%% \vskip 0.5cm
%
\caption{
The singlet-triplet gap $\Delta_{\text{st}}$ of the spin-Peierls model 
as a function of the coupling $g$ for various lattice sizes $N$ for an 
intermediate phonon frequency $J / \omega = 1$. Extrapolations (in 
$1/N$, using the two largest values of $N$) to $N = \infty$ are given by 
the solid diamonds. These are fitted to the K-T form 
$ A f(g) \exp (-B f(g)^2) $, where
$f(g) \equiv \left( g - g_{\text{c}} \right)^{-1/2}$ (solid line). The 
critical coupling $g_{\text{c}}$ is not obtained from this fit. It is 
substantially more accurate to use the gap crossover method (see Table 
\protect\ref{gc_convergence}). The inset shows the extrapolated gap 
(using $N = 32$ and 64) for a small phonon frequency (adiabatic regime) 
$J / \omega = 10$. The dashed line is the result for the static limit 
where the quantum phonon fluctuations are neglected 
\protect\cite{adiabatic}.
\label{gap1}
} 
\end{figure}

For a K-T transition, the gap 
$\Delta = \lim_{N \rightarrow \infty} \Delta_{\text{st}}$
is expected to have an essential singularity at $g = g_{\text{c}}$.
In Fig.\ \ref{gap1}, $\Delta_{\text{st}}$ is plotted as a function
of $g$ for various $N$ in a case of intermediate coupling $J / \omega = 
1$.
Two-point linear extrapolations (in $1/N$) to $N = \infty$ are included
in the plot. These estimates of $\Delta$ are shown to be well fitted by
the K-T form \cite{nomura1}
$\Delta \sim A f(g) \exp( - B (f(g))^2 )$ where
$f(g) \equiv (g - g_{\text{c}})^{-1/2}$. Note that the gap
crossover method (Table \ref{gc_convergence}) is substantially more
accurate than this fitting procedure for determining $g_{\text{c}}$,
the latter tending to overestimate $g_{\text{c}}$ \cite{bursill}.

In the adiabatic regime ($\omega << J$)
there is strong mixing between spin singlet
and phonon excitations. An analogous effect was observed
for the Holstein model \cite{holstein}.
In the case of (\ref{H}) this is manifest in nonlinear corrections
to the scaling of $\Delta_{\text{ss}}$. That is, $\Delta_{\text{ss}}$
is found to be phonon like (flat in $1/N$) until the characteristic
spin energy $2 \pi J / N$ decreases below the bare phonon frequency
$\omega$, at which point $\Delta_{\text{ss}}$ begins to vanish, as $1/N$
$(0 \leq g \leq g_{\text{c}})$, or exponentially
$(g > g_{\text{c}})$. This can be seen in Table \ref{gc_convergence}
from the slow convergence of $g_{\text{c}}(N)$ with $N$ for the
$J / \omega = 10$ case.

Next, we consider the validity of the static approximation in the 
adiabatic regime, where the phonon operators $b_i$ in (1) are replaced 
by the constant dimerization $(-1)^i \delta$, the total energy is 
minimized as a function of $\delta$ then the gap is calculated for this 
optimal value of $\delta$. This calculation was performed by using the 
four-block DMRG method to solve for the ground state energy and gap in 
the dimerised Heisenberg model \cite{adiabatic}. The resulting adiabatic 
curve is compared in Fig.\ \ref{gap1} to the extrapolated gap $\Delta$  
found from our numerical results for $J / \omega = 10$. We see that even 
in this adiabatic region treating the phonons in the mean-field 
approximation is not fully reliable, particularly for the purposes of 
quantitatively fitting the coupling $g$ from the experimental triplet 
gap. The situtation is far worse for phonon frquencies relevant to 
CuGeO$_3$. For example, for the $J / \omega = 1$ case in Fig.\ 
\ref{gap1}, the adiabatic curve would not fit on the same scale as the 
curve from the fully dynamical model.

To consider our results in the context of experiment, estimates of a 
number of parameters for various spin-Peierls compounds are listed in 
Table \ref{experiment}. It can be seen from these estimates and our 
results that the static approximation is highly questionable for 
CuGeO$_3$, and may not be valid for the organic spin-Peierls materials.
A related question is the use of an explicit next-neighbour
($J_2$, frustration) term in adiabatic spin-phonon models of CuGeO$_3$
\cite{fabricuis,castilla}. The value of $J_2$ required to achieve
agreement with susceptibility and magnetic specific heat data is
generally very large ($J_2 / J_1 \approx 0.3)$. Attempts have been made
to justify the inclusion of a $J_2$ term on the basis of Cu-O-O-Cu
superexchange paths \cite{castilla}. However, Ref.\ \onlinecite{uhrig}
and the present analysis suggests that an explicit $J_2$ term
may not be required in order to describe experimental results if the 
phonons are treated quantum mechanically since the phonons induce a 
next-nearest neighbour interaction.

To conclude, we have numerically determined the phase diagram
of a spin-Peierls model (\ref{H}) with high accuracy. Our
results are consistent  with a mapping of the model
to the frustrated spin chain (\ref{j1j2}) in the
anti-adiabatic limit (large phonon frequency).
For a wide range of phonon frequencies compared
to the exchange there is a phase transition
at a non-zero value of the coupling $g$ from
a gapless spin-fluid state to a
gapped dimer phase \cite{sandvik}.
 The transition is in the same universality
class as the Kosterlitz-Thouless transition in 
the frustrated antiferromagnetic chain (\ref{j1j2}). 
Quantum phonon fluctuations are important
in known spin-Peierls materials.

This work was supported by the Australian Research Council. We thank I. 
Affleck, A. Sandvik, O. Sushkov, J. Voit, V. Kotov, H. Fehske, A. Weisse 
and G. Uhrig for discussions and A. Sandvik for showing us results prior 
to publication. Calculations were performed at the New South Wales 
Centre for Parallel Computing and the Australian National University 
Supercomputing Facility.

\begin{table}[h]
\caption{Four-block DMRG convergence of the singlet and triplet
gaps $\Delta_{\text{ss}}$ and $\Delta_{\text{st}}$ of the spin-Peierls
model (\protect\ref{H}) with the truncation parameter $\epsilon$ for
various periodic lattices of size $N$, where $J / \omega = 1$ and
$g / \omega = 0.4$.
}
\begin{tabular}{l|lll}
$N$ & $\epsilon$ & $\Delta_{\text{ss}} / \omega$ &
$\Delta_{\text{st}} / \omega$ \\
\hline
\hline
8     &  $10^{-15}$ &  0.31374961  &  0.5183251 \\
8     &  $10^{-20}$ &  0.31372889  &  0.5183254 \\
8     &  $10^{-22}$ &  0.31372870  &  0.5183254 \\
\hline
%32   &  $10^{-11}$ &  0.0766371   &  0.134382  \\
32    &  $10^{-13}$ &  0.0764782   &  0.133925  \\
32    &  $10^{-15}$ &  0.0765958   &  0.133785  \\
32    &  $10^{-16}$ &  0.0765933   &  0.133778  \\
\hline
128   &  $10^{-10}$ &  0.014909    &  0.04009   \\
128   &  $10^{-11}$ &  0.014817    &  0.03856   \\
%128  &  $10^{-12}$ &  0.014620    &  0.03802   \\
128   &  $10^{-13}$ &  0.014619    &  0.03790   \\
128   &  $10^{-14}$ &  0.014648    &  0.03775   
\end{tabular}
\label{dmrg_convergence}
\end{table}

\begin{table}[h]
\caption{Convergence of the crossover coupling
$g_{\text{c}}(N) / \omega$ with lattice size $N$ for various values of
the adiabaticity parameter $J / \omega$.
$g_{\text{c}}(N)$ is defined by $\Delta_{\text{ss}} = 
\Delta_{\text{st}}$
and converges to the critical coupling $g_{\text{c}}$ as
$N \rightarrow \infty$.
}
\begin{tabular}{c|ccccc}
    & \multicolumn{5}{c}{$J / \omega$} \\
\hline
$N$ &  0.005 &  0.1 &   1.0  &  2.0  &  10.0  \\
\hline
4   &  0.0692 & 0.237 & 0.1201 &  ---  &  ---  \\
8   &  0.0681 & 0.228 & 0.2735 & 0.092 &  ---  \\
16  &  0.0671 & 0.225 & 0.3021 & 0.274 &  ---  \\
32  &   ---   & 0.223 & 0.3087 & 0.310 &  ---  \\
64  &   ---   &  ---  & 0.3092 & 0.318 & 0.249 \\
128 &   ---   &  ---  &   ---  &  ---  & 0.318 \\
256 &   ---   &  ---  &   ---  &  ---  & 0.339 \\
\end{tabular}
\label{gc_convergence}
\end{table}

\begin{table}[htbp]
\caption{
Estimates of the  exchange $J$, phonon frequency $\omega$, 
and energy gap $\Delta$ for various spin-Peierls materials.
All are given in units of Kelvin.
(We are unaware of any other measurements of the frequencies
of the dimerization phonon in organic materials.)
}
\begin{tabular}{l|cccc}
Material & $J$ & $\omega$ & $\Delta$ & Ref.\ \\
\hline
CuGeO$_3$ & 100 & 150, 300 & 20 & \onlinecite{hase,braden}\\
%TTFCuBDT & 77 & 15 & --- & 12 \\
TTFCuS$_4$C$_4$(C$_3$F)$_4$ & 70 & ?\onlinecite{uhrig2} & 20 & 
\onlinecite{bray,kasper}\\
%(TMTTF)$_2$PF$_6$ & 500 & --- & 50 & 17 \\
(MEM)(TCNQ)$_2$ & 50 & 100 & 60 & \onlinecite{huiz,mem}\\
%$\alpha$-NaV$_2$O$_5$ & 400--800 & --- & 90 & 34
%
\end{tabular}
\label{experiment}
\end{table}

\newpage

\begin{references}
\bibitem{bursillnote}
The parameter $\epsilon$, defined in \protect\cite{bursill}, is a 
measure of the amount of Hilbert space that is discarded in the basis 
truncation process.
\bibitem{fitting_note}
Note that for large $N$, the critical coupling is expected to converge 
algebraically with $N$ \protect\cite{castilla}. However, in the regime 
$J/\omega > 1$, the dominant source of the strong $N$ dependence over 
the range of $N$ values studied comes from the mixing of phonon and spin 
excitations \protect\cite{holstein}. Empirically, this contribution to 
the $N$ dependence is found to decay exponentially. We expect a 
crossover to algebraic scaling at large $N$.
\end{references}
\end{document}